\documentclass[twocolumn,showpacs,preprintnumbers,amsmath,amssymb]{revtex4}

\usepackage{graphicx}
\usepackage{dcolumn}
\usepackage{bm}

\newcommand{\eed}{\rm}

\begin{document}

\preprint{APS/123-QED}

\title{Spontaneous Pair Creation revisited}

\author{Pickl, P. }
\email{pickl@mathematik.uni-muenchen.de}
\author{D\"urr, D.}
\email{duerr@mathematik.uni-muenchen.de}


\affiliation{ Mathematisches Institut der Universit\"at M\"unchen\\
Theresienstr. 39\\
80333 M\"unchen\\Germany
}%

\date{\today}

\begin{abstract}
Recently the so called Spontaneous Pair Creation of electron
positron pairs in a strong external field has been rigorously
established. We give here the heuristic core of the proof, since the
results differ from those given in earlier works.
\end{abstract}

\maketitle
\section{Introduction}
The creation of an electron positron pair in an almost stationary very strong external electromagnetic field has
been called---unfortunately misleading---spontaneous pair creation (SPC) (\cite{mprg} - \cite{greiner}). The
phenomenon emerges straight forwardly from the Dirac sea interpretation of negative energy states: An
adiabatically increasing field lifts a particle from the sea to the positive energy subspace where it scatters
and when the potential is gently switched off one has one free electron and one unoccupied state---a hole---in
the sea. The experimental verification has been sought in heavy ion collisions, but without success so far
\cite{exp1}, \cite{exp2}. A rigorous proof of the existence of SPC has been lacking  until recently
\cite{pdneu}. Surprisingly our rigorous proof---the heuristics of which we give here--- yields time estimates
different from what has been reported earlier and which may be relevant for experiments. The rigorous result
concerns the so called external field problem, i.e. interactions between the charges are neglected. Vacuum
polarization will in general perturb the external field (see \cite{hainzl} for a rigorous attempt) and one may
think of the external field as an effective field. The description of SPC in second quantized external field
Dirac theory is known to be equivalent to the existence of certain type of solutions of the one particle Dirac
equation. We shall only discuss the latter.
\section{Formulation of the problem}
Consider the one particle Dirac equation with external electric
potential $A$, a real valued multiple of the $4\times 4$ unit
matrix.  $A mc^2$ gives the potential in the units $eV$.
 On
microscopic timescales $\tau=\frac{mc^2}{\hbar}t$ the equation
reads
\begin{eqnarray}\label{Diracmicalt}
   i\frac{\partial\psi_{\tau}}{\partial \tau}&=&-i\frac{\hbar}{mc}\sum_{l=1}^{3}\alpha_{l}\partial_{l}\psi_{\tau}+A_{\varepsilon \tau}(\mathbf{x})\psi_{\tau}+\beta  \psi_{\tau}
   \nonumber\\&\equiv&(D^0+A_{\varepsilon
   \tau}(\mathbf{x})\psi_{\tau}=D_{\varepsilon \tau}\psi_\tau
\end{eqnarray}
where $\varepsilon$ is a parameter representing the slow time variation of the external potential.

We introduce in (\ref{Diracmicalt}) the macroscopic time scale $s=\varepsilon\tau$:
\begin{eqnarray}\label{Diracmic}
   i\frac{\partial\psi_{s}}{\partial s}\equiv \frac{1}{\varepsilon}D_s\psi_{s}\;.
\end{eqnarray}
%
We will describe pair creation using the Dirac sea interpretation of the Dirac equation. The spectrum of the
Dirac equation without external field is  $(-\infty,-1]\cap[1,\infty)$. Wave functions which lie in the
corresponding positive energy subspace of the free Dirac operator are interpreted as wave functions of particles
or electrons. The Dirac sea  is a many particle wave function built out of wave functions of negative energy
(states) of the free Dirac operator. In the so called vacuum all states in the sea are occupied by particles.
``Holes'' in the Dirac sea are unoccupied states which are interpreted as anti-particles or positrons.

Suppose now that the potential is adiabatically switched on and
later switched off. When the potential is non zero there may be
bound states in the gap, $[-1,1]$. The eigenvalues of these bound
states change with the strength of the potential. The adiabatic
theorem (see e.g. \cite{stefan}) ensures that there is no tunnelling
across spectral gaps. So as long as all bound states are isolated
>from the upper continuous spectrum, transitions from the negative to
the positive energy subspace are adiabatically not possible and the
probability of creating a pair is zero (undercritical case). When
the external field becomes overcritical (at time $s_c=0$), i.e. when
a bound state vanishes in the upper energy subspace, there exists a
solution of the Dirac equation which follows the path of this bound
state from the negative to the positive energy subspace. If the wave
function (or a part of it) does not ``follow'' the bound state back
into the negative energy subspace when the potential is switched off
again, but stays in the positive energy subspace, there will be a
hole in the sea and a freely moving particle: SPC is achieved.
The scenario is symmetric under change of sign of the potential: It
then transports an unoccupied state (a hole) from the positive
energy subspace to the sea and catches a particle  from the sea when
it  is switched off. The hole (positron) then scatters.

To show to what extend the scenario in fact holds one must control
the propagation of the wave function emerging from the bound state
during over-criticality. The wave function will generically not be
a nice scattering state, i.e. its momentum distribution will be
large for small momenta, that is, the wave function lingers around
the range of the potential rather than moving away. We shall
describe now the heuristic core of the situation when  SPC
happens.
\section{The Spectrum of the Dirac operator} We expand the
wave function in generalized eigenfunctions which by themselves
are time dependent. We shall need the eigenfunctions  for times
$\sigma$ close to the critical time. Consider the eigenvalue
equation\eed
\begin{equation}\label{dgmp2}
 E \varphi=D_{\sigma}\varphi
\end{equation}
for fixed $\sigma\in\mathbb{R}$. The continuous subspace determined
by $\mathbf{k}\in \mathbb{R}^3\backslash\{0\}$ or $\mathbf{k}=0$ and
$\sigma\neq0$ is spanned by generalized eigenfunctions (not square
integrable) $\varphi^{j}(\mathbf{k},\sigma,\mathbf{x})$,
$j=1,2,3,4$, with energy $E=\pm E_k=\pm\sqrt{k^2+1}$,
 $j=1,2$ being solutions with positive energy. For ease of notation we
 will drop the spin index $j$ in what follows.

The generalized eigenfunctions also solve the Lippmann Schwinger equation
\begin{eqnarray}\label{LS}
\lefteqn{\varphi
(\sigma,\mathbf{k},\mathbf{x})=\varphi_0(\mathbf{k},\mathbf{x})\nonumber}
\\&&+\int
 G^{+}_{k}(\mathbf{x}-\mathbf{x'}) A_\sigma(\mathbf{x'})
 \varphi(\sigma,\mathbf{k},\mathbf{x'})d^{3}x'\;,
 \end{eqnarray}
with $\varphi_0(\mathbf{k},\mathbf{x})= \xi(\mathbf{k})
e^{i\frac{mc}{\hbar}\mathbf{k}\cdot\mathbf{x}}$, the generalized
eigenfunctions of the free Dirac operator $D^0$, i.e.
  $G^{+}_{k}$ is the kernel of
$(E_{k}-D^0)^{-1}=\lim_{\delta\rightarrow0}(E_{k}-D^0+i\delta)^{-1}$ \cite{thaller}.


Introducing the operator $T_\sigma^k$
\begin{equation}\label{tkdef}
T_\sigma^k f=\int
 G^{+}_{k}(\mathbf{x}-\mathbf{x'}) A_\sigma(\mathbf{x'})
 f(\mathbf{x'})d^{3}x',
 \end{equation}
(\ref{LS})  becomes
\begin{equation}\label{LSE}
(1-
T_\sigma^k)\varphi(\sigma,\mathbf{k},\cdot)=\varphi_0(\mathbf{k},\cdot)\;.
 \end{equation}
There may also exist  resonance-eigenfunctions which are not square integrable but which decay as
$x\rightarrow\infty$.

Finally there may exist eigenfunctions, i.e. bound states
$\Phi_\sigma(\kappa)$ of (\ref{dgmp2}) with energies
$E=\sqrt{1-\kappa^2}\in[-1,1]$.  They satisfy instead of
(\ref{LSE})
\begin{equation}\label{LSEbound}
(1- T_{\sigma(\kappa)}^\kappa)\Phi_{\sigma(\kappa)}=0\;,
 \end{equation} with imaginary $k=i\kappa$.
In the following we will assume that $A$ is such that for each
time $\sigma$ there exist either no or one (degenerate) bound
state $\Phi_\sigma$ and that the bound state vanishes in the
positive continuum at $\sigma=0$. Since there generically exists a
bound state $\Phi$ with energy $1$ (i.e. $\mathbf{k}=0$) at time
$\sigma=0$ and $\partial_\sigma E_\sigma|_{\sigma=0}\neq 0$ (see
\cite{klaus} and our argument below), we will only discuss
that case.

Using (\ref{LSEbound}) for $\sigma=\kappa=0$ observing the
explicit form of $G^+_0$ (see e.g. \cite{duerr}) one can easily
estimate $\Phi$ for large $x$. Assuming $\Phi$ bounded, a term
falling off like $x^{-1}$ appears. Using that this term has to be
equal to zero (so that $\Phi$ is square integrable) one gets the
crucial identity
\begin{equation}\label{klaus}
\int (1+\beta)A_0(\mathbf{x})\Phi(\mathbf{x})d^3x=0\;.
\end{equation}
\subsection{Propagation estimate}
We  estimate the propagation of a wave function generated by the {\em static} Dirac Operator
$D_\sigma=D^0+A_\sigma(\mathbf{x})$, where  $\sigma>0$ should be thought of as near the critical value (the
relevant regime turns out to be of order $\sigma=\mathcal{O}( \epsilon^{1/3})$).

Since the generalized eigenfunctions for $(\sigma,\mathbf{k})\approx(0,0)$ are close to the bound state $\Phi$
it is reasonable to write
 in leading
order:
\begin{equation}\label{dieistrichtig}
\varphi(\sigma,\mathbf{k},\mathbf{x})\approx\eta_\sigma(\mathbf{k})\Phi(\mathbf{x})\;.
\end{equation}
Since they solve (\ref{LS}), the first summand of (\ref{LS}) must
become negligible with respect to $\eta_\sigma(\mathbf{k})\Phi$,
which is part of the second summand. Hence $\eta_\sigma(\mathbf{k})$
must diverge for $(\sigma,\mathbf{k})\to (0,0)$.  For the outgoing
asymptote of the state $\Phi$ (generalized Fourier transform)
evolved with $D_\sigma$ near criticality we have with
(\ref{dieistrichtig}) that
\begin{eqnarray}\label{her}
\widehat{\Phi}_{out}(\sigma,\mathbf{k})&:=&\int
(2\pi)^{-\frac{3}{2}}
\Phi(\mathbf{x})\overline{\varphi}(\sigma,\mathbf{k},\mathbf{x})d3
x\nonumber\\&\approx&(2\pi)^{-\frac{3}{2}}\overline{\eta}_\sigma(\mathbf{k})\;.
\end{eqnarray}
Now, for $(\sigma,k)$ close to but different from $(0,0)$,
$\eta_\sigma(\mathbf{k})\sim\overline{\widehat{\Phi}_{out}}(\sigma,\mathbf{k})$
will be peaked around a value $k(\sigma)$ with width
$\Delta(\sigma)$ (determined below) defined by
\begin{equation}\label{defdelta}
\eta_\sigma(k(\sigma)\pm\Delta(\sigma))\approx\eta_\sigma(k(\sigma))/\sqrt{2}\;.
\end{equation}
We may use the width for the rough estimate
\begin{equation}\label{ableitung}
|\partial_k\widehat{\Phi}_{out}(\sigma,\mathbf{k})|\  <
\Delta(\sigma)^{-1}\widehat{\Phi}_{out}(\sigma,k_\sigma)\;,
\end{equation}
where the right hand side should be multiplied by some appropriate
constant which we---since it is not substantial---take to be one.
 Using
(\ref{dieistrichtig}), (\ref{her}), $d^3k=k^2d\Omega dk$ and
partial integration (observing
$\frac{\varepsilon}{-iks}\partial_ke^{-i(1+\frac{k^2}{2})\frac{s}{\varepsilon}}=e^{-i(1+\frac{k^2}{2})\frac{s}{\varepsilon}}$)
 we get

\begin{eqnarray*}
U_{\sigma}(s,0)\Phi&=&e^{-isD_{\sigma}}\Phi\approx\frac{1}{(2\pi)^{\frac{3}{2}}}\int
e^{-i(1+\frac{k^2}{2})\frac{s}{\varepsilon}}\widehat{\Phi}_{out}\varphi
d3 k
%
%
%
\nonumber\\&=&\frac{-i\varepsilon}{s}
\int e^{-i(1+\frac{k^2}{2})\frac{s}{\varepsilon}}
%
\partial_k\left(|\widehat{\Phi}_{out}|^2\Phi(x)k d\Omega\right) dk
\end{eqnarray*}
By (\ref{ableitung}), (\ref{her}) and (\ref{defdelta}), assuming that $\Delta(\sigma)\ll k(\sigma)$
\begin{eqnarray*}
|\partial_k\left(|\widehat{\Phi}_{out}|^2k\right)|  d\Omega dk&\approx&
|\widehat{\Phi}_{out}|^2\left(\frac{2}{\Delta(\sigma) k}+\frac{1}{k^2}\right) d^3k
\\&\approx&|\widehat{\Phi}_{out}|^2\frac{2}{\Delta(\sigma) k(\sigma)}d^3k
\;.
\end{eqnarray*}
Hence
\begin{eqnarray*}
|U_{\sigma}(s,0)\Phi(\mathbf{x})|&\leq&\frac{2\varepsilon|\Phi(x)|}{s\Delta(\sigma) k(\sigma)}
\int
|\widehat{\Phi}_{out}|^2 d^3k\;.
\end{eqnarray*}
Since $\Phi$ is normalized we get for the decay time $s_d$, defined by $|\langle
U(s_d,0)\Phi,\Phi\rangle|\approx 1/2$
\begin{equation}\label{tautime}
s_d\approx 4\varepsilon (k(\sigma)\Delta(\sigma))^{-1}\;.
\end{equation}
\subsection{Control of the Generalized Eigenfunctions}
Let us now estimate $\eta_\sigma(\mathbf{k})$ for $\sigma \approx 0 $. In view of (\ref{LSE}) and
(\ref{dieistrichtig}) we have that
\begin{equation}\label{LSE2}
(1-T_\sigma^k)\eta_\sigma(\mathbf{k})\Phi\approx\varphi_0(\mathbf{k},\cdot)\;.
 \end{equation}
We can estimate $\eta_\sigma(\mathbf{k})$ by considering the scalar product of (\ref{LSE2}) with $A_0\Phi$:
\begin{eqnarray*}\label{LSE3}
\eta_\sigma(\mathbf{k})\langle(1-
T_\sigma^k)\Phi,A_0\Phi\rangle&\approx&\langle\varphi_0(\mathbf{k},\cdot),A_0\Phi\rangle\;.
 \end{eqnarray*}
Since $\varphi_0(0,\mathbf{x})=\xi(0)$ for which
\begin{equation}\label{phi0}
\xi(0)=D^0\xi(0)=\beta \xi(0)
\end{equation}
we have that $\varphi_0(0,\mathbf{x})=\frac{1+\beta}{2}\xi(0)$ and
in view of (\ref{klaus})
\begin{equation}\label{klaus2}
\langle\varphi_0,A_0\Phi\rangle\mid_{k=0}=\frac{\xi(0)^\dag}{2}\int
(1+\beta)A_0(\mathbf{x})\Phi(\mathbf{x})d^3x=0\;.
\end{equation}
Hence
$\langle\varphi_0(\mathbf{k},\cdot),A_0\Phi\rangle\rangle=\mathcal{O}(k)$,
and it turns out that $\mathcal{O}(k)=C k +\mathcal{O}(k^2)$ with an
appropriate $C\neq0$. Thus
\begin{eqnarray*}
\eta_\sigma(\mathbf{k})\approx Ck\langle (1-T_\sigma^k)\Phi,A_0\Phi\rangle\rangle^{-1}
\end{eqnarray*}
 Expanding $T_\sigma^k$ in orders of $k$ around $k=0$  until fourth
order yields
\begin{eqnarray}\label{taylorsp}
\eta_\sigma(\mathbf{k})\approx
Ck\left(S^\sigma_0+kS^\sigma_1+k^2S^\sigma_2+k^3S^\sigma_3+k^4S^\sigma_4\right)^{-1}\;.
\end{eqnarray}
By virtue of (\ref{tkdef}) one easily shows
\begin{eqnarray}\label{symmetric}
\langle T_\sigma^k\Phi,A_\vartheta\Phi\rangle=\langle A_\sigma\Phi,T_\vartheta^k\Phi\rangle
\end{eqnarray}
and we get with (\ref{LSEbound}), using $A_\sigma>A_0$, that
\begin{eqnarray*}S^\sigma_0&=&\langle(1-T_\sigma0)\Phi,A_0\Phi\rangle=\langle(T0_0-T_\sigma0)\Phi,A_0\Phi\rangle\\&=&\langle
(A_0-A_\sigma)\Phi,T_00\Phi\rangle=\langle
(A_0-A_\sigma)\Phi,\Phi\rangle\\&=&-\parallel\sqrt{A_\sigma-A_0}\Phi\parallel2=-C_0\sigma\;.\end{eqnarray*}
with $C_0>0$. Computing that $\partial_k G_k^+\mid_{k=0}=1+\beta$,
and observing  (\ref{tkdef}) and (\ref{klaus}),
$\partial_kT_k\Phi\mid_{k=0}=0$, hence with (\ref{symmetric})
$$S_1^\sigma=\partial_k\langle T_\sigma^k\Phi,A_0\Phi\rangle\mid_{k=0}=\partial_k\langle T_0^k\Phi,A_\sigma\Phi\rangle\mid_{k=0}=0\;.$$
 Expanding $S^{\sigma}_2$ and $S_3^{\sigma}$ around
$\sigma=0$ we thus obtain
\begin{eqnarray}\label{etareihe} &&\eta_\sigma(\mathbf{k})\\\nonumber&&\approx -Ck\left( C_0\sigma+(C_2+\mathcal{O}(\sigma))
k^{2}+(C_3+\mathcal{O}(\sigma))k^{3}\right)^{-1}\;.
\end{eqnarray}
We shall now determine the constants $C_2$ and $C_3$. For that we
do a similar expansion for (\ref{LSEbound}), i.e. for $\sigma<0$
and for $k=i\kappa$. Dividing (\ref{LSE2}) by
$\eta_\sigma(\mathbf{k})$ and replacing $\Phi$ by
$\Phi_{\sigma(\kappa)}(\approx\Phi)$  yields the left hand side of
(\ref{LSEbound}) and thus $\eta_{\sigma(\kappa)}(i\kappa)=\infty$
(otherwise the right hand side of (\ref{LSEbound}) would not be
zero). Therefore for all $\kappa$
\begin{equation*}
-C_0|\sigma(\kappa)|-(C_2+\mathcal{O}(\sigma(\kappa)))
\kappa^{2}-i(C_3+\mathcal{O}(\sigma(\kappa)))\kappa^{3}+\ldots=0\end{equation*}
Since $C_0 >0$, assuming $C_2,C_3\neq 0$ (see \cite{pdneu} for
a proof) we conclude that
 $C_2<0$ and $C_3$ must be imaginary.  Note that near $k=0$, $\sigma\sim
 k^2=E-m$,  while if $C_1 \neq 0$,  $E-m\sim -\sigma^2$, i.e.
 $dE/d\sigma=0$ at $\sigma=0$, suggesting that a transversal crossing of the
eigenvalue into the upper continuum goes together with a bound
state at $E=1$.

 Hence for (\ref{etareihe}) we get
\begin{eqnarray*}\label{etareihe2}
\eta_\sigma(\mathbf{k})\approx \frac{-Ck}{C_0\sigma-(|C_2|+\mathcal{O}(\sigma))
k^{2}-i(|C_3|+\mathcal{O}(\sigma))k^{3}}\;.
\end{eqnarray*}
For $C_0\sigma\approx C_2k^2$ the denominator behaves like $C_3 k3$, otherwise it behaves like $C_0\sigma-C_2
k^{2}$. Hence by (\ref{her})
\begin{eqnarray}\label{lemgeprobe}
|\widehat{\Phi}_{out}(\sigma,\mathbf{k})|^2\approx Ck^2\left( (C_0\sigma-|C_2| k^{2})^2+|C_3|^2k^{6}\right)^{-1}
\end{eqnarray}
This result \cite{picklneu} differs from the results given in the literature (see e.g. formula (7) in
\cite{mprg}).
%
%
%
The right hand side of (\ref{lemgeprobe}) obviously diverges for
$(\sigma,k)\rightarrow(0,0)$. For fixed $0\neq\sigma\approx 0$ the
divergent behavior is strongest close to the resonance at ($C_0
\sigma- |C_2|k(\sigma)^2=0$)
\begin{equation}\label{knull}
k(\sigma)=\sqrt{\sigma
C_0|C_2|^{-1}}=\mathcal{O}(\sqrt{\sigma})\;.
\end{equation}
In view of (\ref{defdelta}) $\Delta(\sigma)$ can be roughly estimated by setting the right hand side of
(\ref{lemgeprobe}) equal to $1/2$ of its maximal size, i.e
$$C_0\sigma-|C_2| (k(\sigma)+\Delta(\sigma))^{2}\approx |C_3|k^{3}(\sigma)$$
hence
\begin{equation}\label{Delta}\Delta(\sigma)\approx
k(\sigma)^2|C_3| (2|C_2|)^{-1}=\mathcal{O}(\sigma)\;.
\end{equation}
\subsection{Estimating the decay time $s_d$}
Let us first approximate $U$ by $U_{s_d}$, i.e. the time propagator for the time independent Dirac operator
present at $s_d$.

Using (\ref{tautime}), (\ref{knull}) and (\ref{Delta}) we have $s_d\propto\varepsilon s_d^{-\frac{3}{2}}$, hence
$s_d\propto\varepsilon^{\frac{2}{5}}$. The rigorous estimate, taking into account the time dependence of the
external field, yields $s_d\propto\varepsilon^{\frac{1}{3}}$ (\cite{pdneu}). Hence if the field stays
overcritical for times $S\gg\varepsilon^{\frac{1}{3}}$ the probability of pair creation is one. Note that in the
adiabatic case $S=\mathcal{O}(1)$, and thus this is well satisfied.

The distribution of the outgoing momenta will be discussed below.
\section{$\varepsilon$ in Heavy Ion Collisions}
One way to create overcritical fields experimentally are heavy
ion collisions. (There are other experiments, which might become
more relevant for SPC \cite{laser} \cite{neutronen}). Since the
fields of the nuclei are repulsive for positrons, it is the positron
which scatters.
For heavy ion collisions the adiabatic time scale on which the field increases is directly determined by the
relative speed with which the heavy ions approach each other and one computes that $\varepsilon$ is of order
$10^{-1}$, however the time where the field remains overcritical is very small (see \cite{mueller}). The time
variation of the field (even for weak fields)  produces also pairs (see e.g. \cite{bhabba}), which may become
relevant when the field strength is close to criticality and where the time duration of over-criticality is
small (see below). In principle an estimate of the induced pairs is needed  or an experimental measurement of
this ``background'': measuring once a system which is slightly under-critical and compare the rate of created
pairs to a system which is slightly over-critical.

Theoretical models indicate that the collision time can be enlarged if the nuclei form a composite nuclear
system \cite{reinhardt}. This effect may be useful to increase the amount of spontaneously created pairs in U-U
scattering experiments, though it seems not possible to enlarge the collision time beyond the decay time of the
previous chapter.
\subsection{SPC probability for Collision times $\ll s_d$}
In the literature the SPC-probability has been computed for an overcritical field of very short life times like
$S\ll \varepsilon^{\frac{1}{2}}\ll\varepsilon^{\frac{1}{3}}$ \cite{mueller}.

We note that this probability can be easily estimated without any reference to the resonance, contrary to
\cite{PRLgreiner}. We also note, that for such short life times the adiabatically changing field can only vary
by $\mathcal{O}(S)$. However in the literature one considers also the case where the field changes---when it
reaches over critical values---nonadiabatically to a very small overcritical stationary value of the size
$A_0+a$ with $a\ll S$. In this case only a very small part of the critical wave function $\Phi$ will scatter in
the positive continuous subspace. Let $P^\bot$ be the projection onto the subspace orthogonal to $\Phi$. The
probability for scattering may be estimated by $\| P^\bot U(s,0)\Phi\|^2$. Now let
$U_0(s,0)=\exp{-\frac{i}{\epsilon}}s$ then
\begin{eqnarray*}
&&(U(S,0)-U_0(S,0))\Phi\\&&\hspace{1cm}=\frac{i}{\epsilon}\int_0^S
U_0(S,s)(1-D_s)U(s,0)\Phi ds\;.
\end{eqnarray*}
We apply now $P^\bot$ and replace in the integral $U$ by $U_0$, which yields by iteration an error of smaller
order. Observing further that $P^\bot\Phi=0$ and $\Phi=D_0\Phi$
\begin{eqnarray*}
&&\|P^\bot U(S,0)\Phi \|^2\\&&\hspace{1cm}= \|P^\bot\frac{i}{\epsilon}\int_0^S U_0(S,s)(1-D_s)U_0(s,0)\Phi
ds\|^2
\\
&&\hspace{1cm}=\|\frac{i}{\epsilon}e^{\frac{-iS}{\epsilon}}\int_0^S P^\bot(1-D_s)\Phi ds \|^2
\\&&\hspace{1cm}=\|\frac{i}{\epsilon}e^{\frac{-iS}{\epsilon}}\int_0^S P^\bot(A_0-A_s)\Phi ds \|^2
\\&&\hspace{1cm}=\mathcal{O}\left(a\frac{S}{\varepsilon}\right)^2\ll 1
\end{eqnarray*}
%
%
%
This differs from estimates in the literature (see for example
formula (20) in \cite{reinhardt}), where the  formula different from
but corresponding to (\ref{lemgeprobe})(see  e.g. formula 2.24 in
\cite{mueller})  is interpreted as a Breit Wigner form, leading to
false estimates. We note that for the ``short time analysis'' (which is $S\ll s_d$) the use of
Breit-Wigner form is not  reasonable anyhow. 

%
\section{Energy spectrum of the outgoing positrons}
An interesting prediction is the shape of the momentum distribution of the created positron. It would be nice if
the shape would be simply the resonance (\ref{lemgeprobe}) as suggested e.g. in \cite{mueller}. But that
requires a somewhat different situation than what has been discussed here. It would require an overcritical
static field (adiabatic is not enough) of a life time $S$ larger than the decay time $s_d$. Then in fact the
resonance (\ref{lemgeprobe}) would stay more or less intact (see also \cite{reinhardt}). In an adiabatically
changing field the resonance (\ref{lemgeprobe}) changes however with the field and it is highly unclear how. The
resonance (\ref{lemgeprobe}) will be surely washed out and one can only expect a shape which is somewhat peaked
around small momenta.

\end{document}